\documentclass[twocolumn,showpacs,preprintnumbers,amsmath,amssymb,prl]{revtex4}

\usepackage{graphics}
\usepackage{epsfig}
\usepackage{bm}

\def\beq{\begin{equation}}
\def\eeq{\end{equation}}

\def\reff#1{(\ref{#1})}

\def\subsc#1{{\mbox{\rm\scriptsize #1}}}

\def\Wcmcm{\mbox{\rm Wcm$^{-2}$}}

\def\te{t_\mathrm{e}}
\def\tr{t_\mathrm{r}}
\def\Er{E_\mathrm{r}}
\def\pci{{p_{\mathrm{c}i}}}

\def\Pnsdr{P_\mathrm{NSDR}}
\def\Phe{P_\mathrm{He}}
\def\Pheplus{P_{\mathrm{He}^+}}
\def\Pee{P_\mathrm{ee}}

\def\dznsdr{d_z^\mathrm{NSDR}}
\def\dznsdri{d_{zi}^\mathrm{NSDR}}

\def\N3d{N_\subsc{3D}}

\def\Ip{I_\mathrm{p}}
\def\Up{U_\mathrm{p}}

\def\vekt#1{\bm{#1}}

\def\vektr{\vekt{r}}

\def\vekte{\vekt{e}}
\def\vektE{\vekt{E}}

\def\vektA{\vekt{A}}

\def\vektp{\vekt{p}}

\def\vektalpha{\vekt{\alpha}}

\def\operator#1{\hat{#1}}

\def\UopV{\operator{U}^\mathrm{(V)}}

\def\Adach{\hat{A}}
\def\Ahat{\hat{A}}

\def\energy{{\cal{E}}}

\def\bra#1{\langle #1 \vert}
\def\ket#1{| #1 \rangle}

\def\imagi{\mbox{\rm i}}
\def\diff{\,\mbox{\rm d}}

\begin{document}

\title{Nonsequential Double Recombination in Intense Laser Fields}
\date{\today}
\author{P.~Koval}
\author{F.~Wilken}
\author{D.~Bauer}
\email[Corresponding author's e-mail:\ ]{dbauer@mpi-hd.mpg.de}
\author{C.H.~Keitel}
\affiliation{Max-Planck-Institut f\"ur Kernphysik, Postfach 103980, 69029 Heidelberg, Germany}
\date{\today}

\begin{abstract}
A second plateau in the harmonic spectra of laser-driven two-electron atoms is observed both in the numerical solution of a low-dimensional model helium atom and using an extended strong field approximation. It is shown that the harmonics well beyond the usual cut-off are due to the simultaneous recombination of the two electrons, which were emitted during different, previous half-cycles. The new cut-off is explained in terms of classical trajectories. Classical predictions and the time-frequency analysis of the {\em ab initio} quantum results are in excellent agreement. The mechanism corresponds to the inverse single photon double ionization process in the presence of a (low frequency) laser field. 
\end{abstract}


\maketitle

High-order harmonic generation (HOHG) is one of the fundamental processes that occur when intense laser pulses interact with atoms or molecules (see, e.g., Refs.\ \cite{ago04,scr06} for recent reviews).  In the case of a linearly polarized incoming laser field, odd multiples of the  laser frequency are emitted with a relatively high and almost constant efficiency ($\simeq 10^{-6}$) up to the celebrated $\Ip + 3.17\,\Up$ cut-off \cite{lew94}, where $\Ip$ is the ionization potential and $\Up$ is the ponderomotive energy (i.e., the time-averaged quiver energy of a free electron in the laser field). In such a way, coherent short-wavelength radiation down to the ``water window'' could be generated using ``table-top'' equipment \cite{chang97,schnu98}. 

HOHG up to the  $\Ip + 3.17\,\Up$ cut-off is, in a very good approximation, a single active electron-effect, meaning that at a given laser intensity only a single electron, namely the one that is next in the row for sequential ionization, contributes to the harmonic generation.
 Various aspects of two- and many-electron effects on HOHG were studied in Refs.~\cite{lap96,tong01,prag01,band05,zang06,sant06}. However, to the best of our knowledge nonsequential double recombination (NSDR) and the associated second plateau has not been revealed so far.

The known cut-off at $\Ip + 3.17\,\Up$ can be explained within the so-called ``simple man's theory'' (see, e.g., the review Ref.~\cite{milo06} and references therein): an electron is released with vanishing initial velocity at the emission time $\te$ due to ionization by a laser field with a vector potential $\vektA(t)$ and an electric field $\vektE(t)=-\partial_t \vektA(t)$. Thereafter, the electron moves freely in the laser field without being affected anymore by the binding potential $V(\vektr)$, i.e., its momentum and position at time $t>\te$ are given by
$\vektp(t) = \vektA(t) - \vektA(\te)$ and 
$\vektr(t) = \vektalpha(t) -\vektalpha(\te) -\vektA(\te)(t-\te)$, 
respectively,
where $\vektalpha(t)=\int^t\diff t'\,\vektA(t')$ is the excursion of a free electron in the field (atomic units $\hbar, m, \vert e\vert, 4\pi\epsilon_0=1$ are used unless noted otherwise). In order for a harmonic photon being emitted, the electron has to revisit the ion at some time $\tr>\te$, that is, $\vektr(\tr)\stackrel{!}{=} 0$, since only then its overlap with the ground state---and thus the recombination probability---is appreciable. The energy of the emitted harmonic photon is given by $\Omega= \vektp^2(\tr)/2 + \Ip$. Searching the pairs $(\te,\tr)$ for which $\vektr(\tr){=} 0$ {\em and} $\vektp^2(\tr)/2$ is maximum, leads in the case of a constant-amplitude, linearly polarized laser field in dipole approximation, e.g., $\vektA(t)=\hat{A} \vekte_z \sin\omega t$, to $\max(\vektp^2(\tr)/2) = 3.17 \, \Up$ where $\Up=\hat{A}^2/4$. 

\begin{figure}
\includegraphics[width=0.4\textwidth]{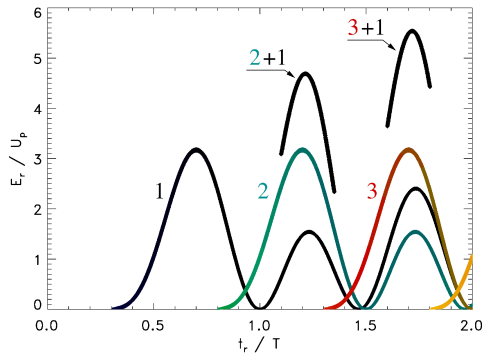}
\caption{(color online). Return energies $\Er$ in units of $\Up$ vs the return time $\tr$ in laser cycles for a laser field $\vektA(t)=\hat{A}\vekte_z \sin\omega t$. Electrons emitted during the first half cycle (1, black), the second  (2, cyan), the third  (3, red) etc.\ reach their maximum return energy $3.17\,\Up$ during the subsequent half cycle. At later returns only lower values are achieved. For two electrons returning at the same time but emitted at different times, the values indicated by ``$2+1$'' and ``$3+1$'' are obtained (the sum ``3+2'' equals ``2+1'' in height and is omitted in the plot).
 \label{retenergs}}
\end{figure}

Let us now consider a two-electron atom or ion where the two electrons are freed by sequential ionization. The possible kinetic return energies $\Er=\vektp^2(\tr)/2$ are shown in Fig.~\ref{retenergs}. Each of the two electrons has a maximum return energy $3.17\,\Up$. If both electrons moved along the same trajectory in the continuum and recombined together, emitting a {\em single} photon, one would expect a HOHG cut-off at $2\cdot 3.17\,\Up + I_\mathrm{p}^{(1)} + I_\mathrm{p}^{(2)}$ with $I_{\mathrm{p}}^{(i)}$ the ionization potential for the $i$th electron. However, electron repulsion renders this process extremely unlikely since the two electrons would have to ``sit on top of each other'' for about half a laser cycle (which is even less likely than ``collective tunneling'' \cite{eichm00}). Instead, if the electrons are emitted at certain times during subsequent or next but one half cycles, there are trajectories that do not cross before the recombination event, so that their mutual repulsion plays only a minor role (and is compensated by the attraction of the ionic potential). If, with respect to the first electron, the second electron is emitted during the subsequent half cycle, the sum of return energies around its maximum value is shown in Fig.~\ref{retenergs}  (indicated by ``$2+1$''). For the constant-amplitude pulse one finds $4.70\,\Up$. If, however, the delay between the emission of first and second electron is greater than half a cycle the sum of the return energies can be even higher, as clearly visible from the values labelled by ``$3+1$'' in Fig.~\ref{retenergs}. The HOHG cut-off for NSDR is then expected at
\[ \max(\Omega_\mathrm{NSDR}) = 5.55\,\Up +  I_\mathrm{p}^{(1)} + I_\mathrm{p}^{(2)}. \] 
In the following, we will show that this cut-off can be clearly identified in {\em ab initio} solutions of the two-electron, time-dependent Schr\"odinger equation (TDSE) and using an extended strong field approximation (SFA).  

First, we employ a widely used one-dimensional (1D) model He-atom \cite{1dhe} to study NSDR on an {\em ab initio} TDSE level%
. 
The Hamiltonian in dipole approximation reads
\[ \hat{H}(t) = \sum_{i=1}^2 \left( \frac{[\pci+A(t)]^2}{2} + V(x_i)\right) + W(x_1-x_2), \]
with $V(x_i) = - {Z} (x_i^2+\epsilon_\mathrm{ei})^{-1/2}$, $W(x_1-x_2)=[(x_1-x_2)^2+\epsilon_\mathrm{ee}]^{-1/2}$, $Z=2$, and $\pci$ the canonical momenta of the two electrons, i.e., $\dot{x}_i=\pci+A(t)$. The soft-core parameters $\epsilon_\mathrm{ei}$, $\epsilon_\mathrm{ee}$ can be tuned in such a way that the model ionization potentials $ I_\mathrm{p}^{(i)}$ equal the real ones of the 3D He atom. The actual values were $\epsilon_\mathrm{ei}=0.5$ (which yields the correct $ I_\mathrm{p}^{(2)}=2.0$ for He$^+$) and $\epsilon_\mathrm{ee}=0.329$ (so that $ I_\mathrm{p}^{(1)}=0.904$) \cite{remarkI}.
 The TDSE $\imagi\partial_t \Psi(x_1,x_2,t) = \hat{H}(t) \Psi(x_1,x_2,t)$ was solved on a $x_1,x_2$-grid ($\Delta x=0.2$) using a split-operator Crank-Nicolson (Peaceman-Rachford) approach ($\Delta t=0.075$), starting at $t=0$ with the spatially symmetric spin-singlet ground state wave function. The $n$-cycle laser pulse was of the form $A(t)=\Adach \sin^2(\omega t/2n) \sin\omega t$ for $0\leq t \leq nT$ and zero otherwise (with $T=2\pi/\omega$). The harmonic spectra are calculated from the modulus-square of the Fourier-transformed acceleration $a(t)$ \cite{hohgcalc}.

\begin{figure}
\includegraphics[width=0.4\textwidth]{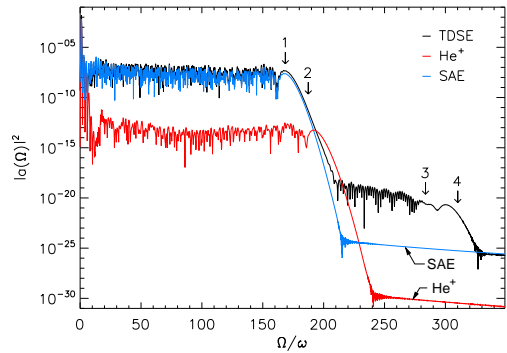}
  \caption{(color online). HOHG spectra for a $n=6$-cycle laser pulse with $\omega=0.0584$ and $\Adach=3.417$ ($I=1.4\cdot 10^{15}$\,\Wcmcm). The spectrum for the He-model with both electrons active (drawn black) shows a second plateau. The corresponding single active electron (SAE)-result  and the spectrum obtained from the He$^+$-ion  are also included. The arrows indicate the expected cut-off positions (see text). \label{specI}}
\end{figure}

Results for HOHG spectra obtained from the He model atom exposed to a $n=6$-cycle laser pulse with $\omega=0.0584$ and $\Adach=3.417$ ($I=1.4\cdot 10^{15}$\,\Wcmcm) are shown in Fig.~\ref{specI}. Besides the result for the fully correlated system with both electrons active, the corresponding spectra for a frozen, inner electron (labelled ``SAE'', drawn blue) and He$^+$ (drawn red) are shown.
For the single active electron (SAE)-calculation, a frozen Hartree-Fock potential (leading to the proper ionization potential for the outer electron $I_\mathrm{p}^{(1)}=0.904$) was used. Not surprisingly, the SAE spectrum agrees well with the two active electron (TAE)-result throughout the ``usual'' plateau (whose cut-off $3.17\,\Up+I_\mathrm{p}^{(1)}$ \cite{remarkII} is indicated by arrow 1) because at the chosen laser intensity it is the outer electron that contributes most to HOHG. As expected, the plateau of the He$^+$-spectrum is several orders of magnitude lower in efficiency. The He$^+$ HOHG-cut-off is expected at  $3.17\,\Up+I_\mathrm{p}^{(2)}$ (indicated by arrow 2). Neither the SAE- nor the  He$^+$-spectrum show a second plateau. The TAE-result, however, confirms our above considerations. Arrows 3 and 4 indicate the positions $4.70\,\Up +  I_\mathrm{p}^{(1)} + I_\mathrm{p}^{(2)}$ and $5.55\,\Up +  I_\mathrm{p}^{(1)} + I_\mathrm{p}^{(2)}$, respectively \cite{remarkII}. 
Around arrow 3 a qualitative change occurs: for lower harmonic frequencies the spectrum displays a rich interference structure because many  ``quantum trajectories'' \cite{milo06} contribute. The situation changes for the harmonic orders where only the classical solutions with $\tr/T\in[3,3.5]$ survive. There are only two such solutions (and finally, at the cut-off, only one), resulting in a much less jagged spectrum between arrows 3 and 4.

Switching-off electron-electron correlation in the two-electron  TDSE-code is equivalent to the simulation of two independent He$^+$-ions. For the latter, NSDR is impossible, showing that electron correlation is clearly required for NSDR to happen since otherwise the two electrons cannot emit their total energy in a {\em single} photon. This is well known for the inverse process, i.e., single photon double ionization.

\begin{figure}
\includegraphics[width=0.4\textwidth]{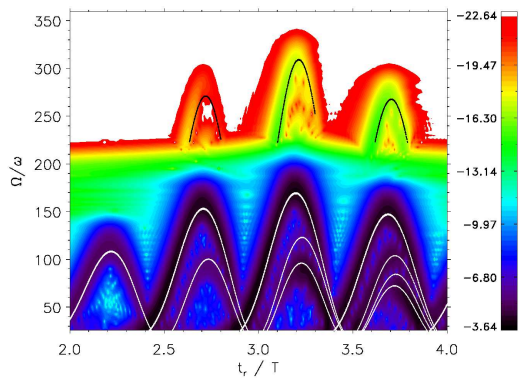}
\caption{Time-frequency analysis of the HOHG spectrum in Fig.~\ref{specI}, showing the color-coded contour plot of $\log_{10} \vert a(\Omega,t) \vert^2$. The relevant classical solutions for the 6-cycle $\sin^2$-pulse are superimposed. The possible classical recombination energies of a single electron $\Omega = \Er + \Ip^{(1)}$ are drawn white. The sums of the two highest classical return energies $ E_{\mathrm{r}1}$,  $ E_{\mathrm{r}2}$ plus the two ionization potentials, $\Omega=\sum_i [ E_{\mathrm{r}i} + \Ip^{(i)}]$,  are superimposed in black for the recombination times $\tr$ of interest.   \label{wt}  }
\end{figure}

The classical trajectories directly emerge in the TDSE quantum results if a time-frequency analysis is performed. To that end a window is applied to the spectrum, selecting a certain frequency interval. The result is Fourier-transformed back, leading to a complex quantity $a(\Omega,t)$ containing information about when the selected harmonics are emitted. 
The time-frequency analysis of the TAE-spectrum of Fig.~\ref{specI} is shown in Fig.~\ref{wt}. The relevant classical ``simple man's'' solutions are superimposed (analogous to Fig.~\ref{retenergs}). The recombination energies $\Er + \Ip^{(1)}$ of individual electrons are drawn white. In the frequency intervals of interest, the sum of the two classical highest return energies (plus the ionization potentials) are superimposed in black. The excellent agreement between ``simple man's'' and {\em ab initio} quantum results shows that NSDR is indeed the mechanism behind the second plateau.

The instants of ionization of the two electrons can be controlled by applying attosecond pulses of photon energies close to the ionization potentials. In that way we were able to enhance the strength of the NSDR plateau in Fig.~\ref{specI}  by three orders of magnitude. More details will be given in a forthcoming publication.

One may object that a 1D model He overestimates the efficiency of the NSDR process because of a reduced wave-packet spreading while the electrons are in the continuum, or other effects. The remainder of this Letter is therefore devoted to NSDR from an SFA perspective where the process can be studied in full dimensionality. Within the SFA the dipole in polarization direction $\dznsdr(t)$ responsible for HOHG via NSDR can be derived similarly to the case of nonsequential double ionization \cite{nsdi} and reads $\dznsdr(t) = \sum_{i=1}^2 \dznsdri(t)$ with
\begin{widetext}
\begin{eqnarray}
\dznsdri(t) & =& \imagi \int_0^t\!\!\diff t' \int_0^{t'}\!\!\diff t_1 \int_{0}^{t'}\!\!\diff t_2\, \bra{\Phi_1\Phi_2} V_{12}(t') \UopV_1\otimes \UopV_2(t',t) z_i\UopV_1\otimes \UopV_2(t,t_2) E(t_2)z_2 \label{dznsdr}\\
&& \qquad\qquad\qquad\qquad\qquad\times \UopV_1(t_2,t_1)   E(t_1)z_1 \exp[-\imagi \energy_1(t_1-t')] \exp[-\imagi \energy_2(t_2-t')]  \ket{\Phi_1\Phi_2} + \mathrm{h.c.} \nonumber
\end{eqnarray}
\end{widetext}
The interpretation of $\dznsdr(t)$ is as follows: electrons 1 and 2 start both in their ground states $\ket{\Phi_1}$, $\ket{\Phi_2}$ with binding energies $\energy_1$, $\energy_2$, are dislodged by the electric field $E$ at times $t_1$ and $t_2$, respectively, interact at time $t'\geq t_1,t_2$ via $V_{12}$, and recombine at time $t\geq t'$. The Volkov time evolution operator $\UopV_i$, $i=1,2$ governs the propagation of electron $i$ in the laser field (without any Coulomb interaction). We evaluate \reff{dznsdr} using hydrogenic ground states with energies $\energy_1=-0.9$ and $\energy_2=-2.0$, the expansion of the length gauge Volkov-propagators in Volkov-waves (see, e.g.~\cite{milo06}), assuming a contact-type interaction $V_{12}=\delta(\vektr_1)\delta(\vektr_1-\vektr_2)$ \cite{nsdi}, and employing the standard saddle-point integration over intermediate electron momenta \cite{lew94}. In the case of a contact interaction, the triple time integral in the final expressions for  the $\dznsdri(t)$ can be disentangled into a double time integral over the interaction time $t'$, the ionization time $t_i$, and an integral over the remaining ionization time $t_{j\neq i}$ that needs to be calculated only once for all $t'$, substantially reducing the numerical effort.
\begin{figure}
\includegraphics[width=0.4\textwidth]{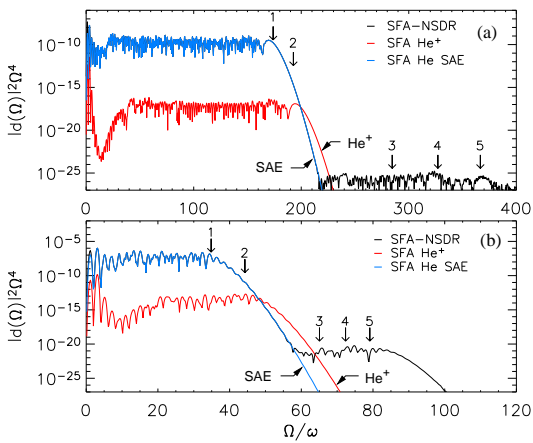}
\caption{(color online). High-order harmonic spectra obtained by Fourier-transforming the SFA-NSDR expression for the dipole $\dznsdr(t)$ (black).  For comparison, the single active electron-result for He (blue) and He$^+$ (red) are included; (a) same laser parameters as in  Fig.~\ref{specI}; (b) $\omega=2\cdot 0.0584$, $\hat{A}=2.0$. Classically expected cut-offs are indicated by arrows: (1) $\vert\energy_{1}\vert+ 3.17 \Up$, (2) $\vert\energy_{2}\vert+ 3.17 \Up$, (3) $\vert\energy_{1}\vert+\vert\energy_{2}\vert +  4.70 \Up$,  (4) $\vert\energy_{1}\vert+\vert \energy_{2}\vert+  5.55 \Up$, and (5) $\vert\energy_{1}\vert+\vert\energy_{2}\vert + 2\cdot 3.17 \Up$. \label{sfaspectra} }
\end{figure}

Figure~\ref{sfaspectra} shows spectra obtained by Fourier-transforming $\dznsdr(t)$ for (a) the same laser parameters as in Fig.~\ref{specI} and (b) a six-cycle pulse of twice the frequency and $\hat{A}=2.0$. A second plateau is obtained in both cases although with very different efficiencies. The SFA result for  $\vert \dznsdr(\Omega) \vert^2\Omega^4$ is, owing to the wave packet spreading of the two electrons,  proportional to $\omega^2$ and $\omega^6$ in one and three dimensions, respectively. As a result, the level of the NSDR plateau in the low frequency case of Fig.~\ref{sfaspectra}a is much lower than in Fig.~\ref{specI}. The wave packet spreading is substantially reduced for the doubled frequency, leading to much more efficient NSDR in  Fig.~\ref{sfaspectra}b.  The probability $\Pnsdr$ of NSDR to occur is $\Pnsdr=\Phe \Pheplus\Pee$ where $\Phe$, $\Pheplus$ are the probabilities for HOHG from He, and He$^+$, respectively, and $\Pee$ is the probability for the emission of a single photon due to electron-correlation (instead of two lower-energy photons). Because in NSDR both electrons return to the nucleus at the same time, electron-electron interaction is very likely, i.e., $\Pee\simeq 1$ is expected, and confirmed by both the 1D TDSE and 3D SFA results.
The level of the NSDR plateau can thus be estimated by simply multiplying the levels of the SAE He and He$^+$ plateaus. 
The unlikely events of simultaneous tunneling ionization ($t_1=t_2$) and motion of the two electrons along the same trajectory are not inhibited in the SFA dipole \reff{dznsdr}.   As a consequence, the SFA-NSDR spectrum extends up to $\vert\energy_{1}\vert+\vert\energy_{2}\vert + 2\cdot 3.17 \Up$, a fact which, however, is irrelevant  for the determination of the relative strength of the second plateau.

In conclusion, the existence of nonsequential double recombination in intense laser fields 
was revealed. In this process, two electrons are freed by the laser, move in the continuum, and are driven back to the ion by the laser, where they recombine together upon emission of their joint energy plus their ionization potentials as a {\em single} photon. 
Nonsequential double recombination may thus be viewed as inverse single photon double ionization. It manifests itself in a second plateau in high-order harmonic generation spectra. This was demonstrated both by solving the time-dependent Schr\"odinger equation of a two-electron model atom numerically and using the strong field approximation. The efficiency of the process is approximately given by the product of the efficiencies for high-order harmonic generation in He and He$^+$. It may be substantially enhanced by aiding the desired ionization times with the help of xuv attosecond pulses.    The main features of the harmonic spectra, including the position of the new cut-off,  were explained in terms of classical trajectories. 
Finally, it should be noted that, instead of simultaneous recombination, the two returning electrons may undergo elastic or inelastic $(2e,ne)$-scattering processes such as, e.g., nonsequential ionization involving $2+n$ electrons or  higher-order above-threshold ionization.

This work was supported by the Deutsche Forschungsgemeinschaft.



\begin{thebibliography}{}
\bibitem{remarkI} One could also set $\epsilon_\mathrm{ee}=\epsilon_\mathrm{ei}$ and use $Z$ as the other free parameter. The results presented in this Letter are insensitive to such details of the model.
\bibitem{remarkII} Note that in few-cycle pulses the actual vector potential $A(t)$ may never reach $\Ahat$. As a consequence, the $\Up$ to be used in the cut-off laws has to be adapted.
\end{thebibliography}
\end{document}